\documentclass[preprintnumbers,prd,onecolumn,floatfix,superscriptaddress, nofootinbib]{revtex4-2}

\usepackage{setspace}
\usepackage{graphicx}
\usepackage{subfig}
\usepackage{epsfig}
\usepackage{bm}
\usepackage{amssymb}
\usepackage{float}
\usepackage{amsmath}
\usepackage{dcolumn}
\usepackage[colorlinks]{hyperref}
\usepackage[usenames,dvipsnames]{color}
\usepackage{enumitem}

\hypersetup{ breaklinks=true, pdfstartview={FitH}, colorlinks=true, linkcolor=blue, citecolor=red, filecolor=magenta, urlcolor=blue, anchorcolor=green, linktocpage=true }

\def\doi{http://doi.org}

\newcommand{\be}{\begin{equation}}
\newcommand{\ee}{\end{equation}}
\newcommand{\beano}{\begin{eqnarray*}}
\newcommand{\eeano}{\end{eqnarray*}}
\newcommand{\ba}{\begin{eqnarray}}
\newcommand{\ea}{\end{eqnarray}}

\begin{document}

\title{ FLRW cosmology in Weyl type $f(Q)$ gravity and observational constraints}
\author{G. K. Goswami}
\email{gk.goswami9@gmail.com}
\affiliation{Department of Mathematics, Netaji Subhas University of Technology, New Delhi-110 078, India}
\author{Rita Rani}
\email{ritarani508@gmail.com}
\affiliation{Department of Mathematics, Netaji Subhas University of Technology, New Delhi-110 078, India}
\author{J. K. Singh}
\email{jksingh@nsut.ac.in}
\affiliation{Department of Mathematics, Netaji Subhas University of Technology, New Delhi-110 078, India}
\author{Anirudh Pradhan}
\email{pradhan.anirudh@gmail.com}
\affiliation{Centre for Cosmology, Astrophysics and Space Science (CCASS), GLA University, Mathura-281 406, Uttar Pradesh, India}

\begin{abstract}
\begin{singlespace}
We propose to develop a cosmological model of the universe based on Weyl type $ f(Q) $ gravity which shows the transition from decelerating in the past to acceleration at present by considering a particular functional form of  $ f(Q) $ gravity as $ f(Q) = ({H_0}^2) (\alpha_1 + \alpha_2 \hskip0.05in log ({H_0^{-2}} Q)) $. We have solved Weyl type $ f(Q) $ gravity field equations numerically and have obtained numerical solutions to the Hubble and deceleration parameters, distance modulus, and apparent magnitudes of stellar objects like SNIa Supernovae. We have also obtained numerical solutions for the Weyl vector $ w $, non-metricity scalar $ Q $, and the Lagrangian multiplier $ \lambda $ appearing in the action of $ f(Q) $ gravity. We have compared our theoretical solutions with the error bar plots of the Observed Hubble data set of $ 77 $ points, $ 580 $ distance modulus SNIa data set, and $ 1048 $ supernova Pantheon data sets of apparent magnitudes. It is found that our results fit well with the observed data set points. The model envisages a unique feature that although the universe is filled with perfect fluid as dust whose pressure is zero, the weyl vector dominance $ f(Q) $ creates acceleration in it.
 
\end{singlespace}
\end{abstract}

\maketitle


\section{Introduction}{\label{sec-1}}
In the year 1915, Einstein completely replaced the instantaneous action at a distance nature of gravitation with a field theory of general relativity (GR) \cite{Einstein:1915, Einstein:1922}. Gravitation was geometrized due to its permanent nature. The uniform distribution of gravitational structures in the universe over cosmic range makes it a spatially homogeneous and isotropic 4-dimensional space-time of constant curvature. These were the novel ideas of GR. Long back before Einstein, Riemann \cite{Reiman:1919} developed the geometry of higher dimensional curved spaces with the help of tensor algebra and calculus. It includes the space-time that consists of metric and affine structures that are determined by metric tensor $ g_{i j} $ and Christoffel symbol $ \Gamma^\alpha_{i j} $. Einstein used Riemannian geometry as a mathematical tool to describe the curved space-time generated by the gravitational field in the universe. The four crucial tests of GR and the FLRW cosmological model that gives initial unavoidable big bang singularity tell the success story of GR. 
In the last few decades, the study indicates that the universe is expanding and accelerating. It is confirmed by cosmological observations such as Type Ia supernovae \cite{SupernovaSearchTeam:1998fmf, SupernovaCosmologyProject:1998vns}, cosmic microwave observations \cite{WMAP:2010qai} and Planck data \cite{Planck:2018vyg}. Scientists have modified general relativity in various ways to support that the universe is expanding and accelerating. Some of the modified theories include $ f(R) $ where $ R $ is the Ricci scalar \cite{Nojiri:2003ft, Starobinsky:2007hu, Sotiriou:2006mu, Sotiriou:2006qn, Srivastava:2006xq, Sporea:2014apa, Nojiri:2022ski, Goswami:2022vfq}, $ f(R,T) $ an extension of $ f(R) $ gravity with the trace ($ T $) of energy-momentum tensor \cite{Harko:2011kv, Singh:2018xjv, Goswami:2019zci, Singh:2020gxd, Bhardwaj:2022lrm, Singh:2022nfm, Singh:2022eun, Singh:2022jhep, Pradhan:2022lne, Pradhan:2023mku, Pradhan:2023nha}, $ f(G) $ where $ G $ is the Gauss-Bonnet Tensor \cite{DeFelice:2008wz, Sharif:2016drh, MontelongoGarcia:2010ip} and $ f(R, G) $ gravity \cite{Odintsov:2018nch, DeLaurentis:2015fea}.

In the year 1916, German mathematician Hermann Weyl \cite{Weyl:1918} proposed an extension of Riemannian geometry that unified the theory of gravity and electromagnetism. Weyl introduced an intrinsic vector field $w^\alpha$ and a semi-metric connection  ${\tilde{\Gamma}^\alpha}_{i j}$ to define parallel transportation of a vector from one point to another in such a way that both its direction and magnitude change. However, it faced withdrawal due to Einstein's criticism of the theory. After this, an extension of general relativity was proposed by Cartan in which he introduced a Torsion field \cite{Cartan:1922}.
This led to the new extension of general relativity known as the Einstein-Cartan theory
\cite{Cartan:1923, Cartan:1924, Cartan:1925, Hehl:1976}. Same time, Weitsenb$\ddot{o}$ck introduced a theory based on Weitsenb$\ddot{o}$ck space with torsion and zero Reimann curvature \cite{Weitsenbock:1923}. The idea leads to the concept of distant parallelism which is known as teleparallelism or absolute parallelism. The primary idea used in the teleparallel formulation of gravity is to use tetrad vectors instead of metric $ g_{ij} $ of the spacetime that describes the gravitational phenomenon. This led to the concept of the teleparallel equivalent of General Relativity (TEGR) \cite{Hayashi:1979qx}. So in the following years, Scientists like Dirac, Cartan, Weitezenb$\ddot{o}$ck, and many more started working on Weyl geometry-based spaces and have proposed the extension to the Weyl gravity such as Weyl-Dirac Langragian \cite{Dirac:1973, Dirac:1974, Rosen:1982nr, Israelit:2010jc}, Weyl-Cartan theory \cite{HHvon:1997, Puetzfeld:2001hk}, Weyl-Cartan-Weitzenb$\ddot{o}$ck theory \cite{Haghani:2012bt, Haghani:2013pea}.

In fact, there are two geometric equivalent frameworks of  Riemannian geometry. The one is the teleparallel framework in which the curvature and the nonmetricity are zero i.e. it is entirely based on the torsion. The second one is the geometry that is completely described by the nonmetricity ($ Q $) which is known as symmetric teleparallel gravity \cite{Nester:1998mp}.

The symmetric teleparallel gravity was further extended into   $ f(Q) $ theory  \cite{BeltranJimenez:2017tkd}. Beltran et al. \cite{BeltranJimenez:2019tme} studied the concept of cosmological implications in $ f(Q) $ gravity. 
Mandal et al. \cite{Mandal:2020buf, Mandal:2020lyq} analyzed the cosmography in $ f(Q) $ gravity and discussed the energy conditions of $ f(Q) $ cosmology respectively. 
W. Khyllep et al. \cite{Khyllep:2021pcu} investigate the cosmological behavior at the background and perturbation level of the power-law model of $ f(Q) $ theory. Off late, Kun Hu et al.\cite{ Hu:2023ndc} constructed the bounce inflation model for the early universe, and calculated the tensor perturbations (namely, primordial gravitational waves) of the model. Many others recent works in $ f(Q) $ gravity include \cite{Dimakis:2021gby, Frusciante:2021sio, Lin:2021uqa, Capozziello:2022zzh, DAgostino:2022tdk, Calza:2022mwt}.

We propose to develop a cosmological model of the universe based on Weyl type $ f(Q) $ gravity which carries a salient feature that in the past the universe was decelerating. After a certain epoch, it starts accelerating and still continuing at present. For this, the particular functional form of  $ f(Q) $ gravity is taken as $ f(Q) = \alpha_1 + \alpha_2 log ({H_0^{-2}}) Q $. We have solved numerically the  Weyl type $ f(Q) $ gravity field equations and have obtained numerical solutions to the Hubble and deceleration parameters, distance modulus, and apparent magnitudes of stellar objects like SNIa Supernovae. We have also obtained numerical solutions for the Weyl vector, non-metricity scalar, and the Lagrangian multiplier $ \lambda $ appearing in the action of $ f(Q) $ gravity. We have compared our theoretical solutions with the error bar plots of the Observed Hubble data set of $ 77 $ points, $ 580 $ distance modulus SNIa data set, and $ 1048 $ supernova Pantheon data sets of apparent magnitudes. It is found that our results fit well with the observed data set points. The model envisages a unique feature that although the universe is filled with perfect fluid as dust whose pressure is zero, the weyl vector dominance $ f(Q) $ creates acceleration in it.

The paper is structured as follows. In Sec. \ref{sec-2}, we have presented Weyl type $ f(Q) $ gravity action and field equations. In Sec. \ref{sec-3}, we solve the field equations for FLRW space-time by taking the energy-momentum tensor as that of a perfect fluid and obtained numerical solutions to the Hubble and deceleration parameters, distance modulus, and apparent magnitudes of stellar objects like SNIa Supernovae. We have also obtained numerical solutions for the Weyl vector $ w $ and the Lagrangian multiplier $ \lambda $ appearing in the action of $ f(Q) $ gravity. In this section, we have also compared the cosmological parameters with the standard $ \Lambda $CDM. In Sec. \ref{sec-4}, we compare our theoretical solutions with the Observed Hubble data set of $ 77 $ points, $ 580 $ distance modulus SNIa data set, and $ 1048 $ supernova Pantheon data sets of apparent magnitudes. Finally in the last Sec. \ref{sec-5} we have concluded the work.


\section{ Field Equations of the Weyl type $ f(Q) $ theory}{\label{sec-2}}

The action in Weyl-type $ f(Q) $ gravity is given by \cite{Xu:2020yeg}

\begin{equation}{\label{1}}
     S = \int \bigg[ k^2 f(Q) - \frac{1}{4} W_{i j} W^{i j} - \frac{1}{2} m^2 w_{i} w^{i} + \lambda (R+ 6 \nabla_{\alpha} w^{\alpha} -6 w_{\alpha} w^{\alpha} ) + L_m \bigg] \sqrt{-g}  d^4 x 
\end{equation}

where $ k^2 \equiv \frac{1}{16 \pi G} $, $ m $ is the mass of the particle associated with the intrinsic vector field $ w_{i} $ of Weyl geometry, $ L_m $ is the matter Lagrangian and $ f(Q) $ is a general function of non-metricity scalar $ Q $. The second and third term represents the ordinary kinetic term and mass term of the vector field respectively. The Lagrangian multiplier scalar $\lambda$ is put to make the Weyl geometry a curved space-time. The brief introduction to Weyl geometry which introduces the intrinsic vector field $ w_{i} $, non-metricity scalar $ Q $ and the tensor $ W^{i j} $ is described in the Appendix.   

We obtain the following Proca type equation by varying the action (\ref{1}) with respect to the vector field $ w $,

\begin{equation}{\label{2}}
     \nabla^j W_{i j} - (m^2 + 12 k^2 f_Q + 12 \lambda) w_i = 6 \nabla_i \lambda.
\end{equation}

If we compare the Eq. (\ref{2}) with the standard Proca equation, we may define an effective dynamical mass of the vector field as follows

\begin{equation}{\label{3}}
     m^2_{eff} = m^2 + 12 k^2 f_Q + 12 \lambda
\end{equation}
 
By varying the action (\ref{2}) with respect to the metric, we obtain the field equation,

\begin{multline}{\label{4}}
 \frac{1}{2} (T_{i j} + S_{i j} ) = - \frac{k^2}{2} g_{i j}  f -6 k^2 f_Q w_{i} w_{j} + \lambda ( R_{i j} - 6 w_{i} w_{j} + 3 g_{i j} \nabla_\gamma w^{\gamma}) +\\
  3 g_{i j} w ^{\gamma} \nabla_{\gamma} \lambda - 6 w_{(i \nabla_j)} \lambda + g_{i j}  \square \lambda - \nabla_{j} \nabla_{i} \lambda,
\end{multline}

where $ f_Q $ is the derivative of $ f $ with respect $ Q $, $ T_{ij} $ is the energy-momentum tensor of the content of the universe,

\begin{equation}{\label{5}}
    T_{i j}  \equiv -\frac{2}{\sqrt{-g}} \frac{\delta(\sqrt{-g} L_m)}{\delta g^{i j}} 
\end{equation}
 
and $ S_{i j} $ represents the re-scaled energy-momentum tensor of the free Proca field,
 
\begin{equation}{\label{6}}
    S_{i j} = -\frac{1}{4} g_{i j}  W_{\eta \alpha} W^{\eta \alpha} + W_{i \eta} W_{j}^{\eta} -\frac{1}{2} m^2 g_{i j} w_{\eta} w^{\eta} + m^2 w_{i} w_{j}.
\end{equation}


\section{Cosmological Evolution in flat FLRW metric}{\label{sec-3}}
We consider the following spatially flat Friedmann-Lemaitre-Robertson-Walker (FLRW) metric which describes the cosmological evolution in a flat geometry,

\begin{equation}{\label{7}}
   ds^2 = -dt^2 + a^2(t)(dx^2+dy^2+dz^2).
\end{equation}

where $ a(t) $ is a scale factor. The vector field $ w_{i} $ is taken as  as $  w_{i} = [0,0,0,\psi(t)] $. Therefore, $ w^2 = w_i w^{i} = - \psi^2(t) $ and $ Q = -6 w^2 = 6 \psi^2(t) $. The Lagrangian of the perfect fluid is taken as $ L_m = p $. Therefore,

\begin{equation}{\label{8}}
    T^i_j = (p + \rho)u^i u_j + p \delta^i_j = diag( p, p, p, -\rho),
\end{equation}

where $ p $ and $ \rho $ are the pressure and matter-energy density of the perfect fluid. We have considered velocity vector $ u^i = (0,0,0,1) $, so that $ u^i u_i = -1 $.

The generalized Proca equation for metric Eq.(\ref{7}) can be written as, 
 \begin{align}
    \dot{\psi} &= \dot{H} + 2 H^2 + \psi^2 - 3 H \psi, \label{9}\\
     \dot{\lambda} &= (-\frac{1}{6} m^2 - 2 k^2 f_Q -2 \lambda ) \psi = - \frac{1}{6} m_{eff}^2 \psi,  \label{10} \\
     \partial_i \lambda &= 0.\label{11}
 \end{align}

The field equations Eq.(\ref{4}) for metric Eq.(\ref{7}) are obtained as, 
\begin{align}
    \frac{1}{2} \rho &= \frac{k^2}{2} f - \bigg( 6 k^2 f_Q + \frac{1}{4} m^2 \bigg) \psi^2 
      - 3 \lambda (\psi^2 - H^2 ) - 3 \dot{\lambda} (\psi - H),{\label{12}} \\
      -\frac{1}{2} p &= \frac{k^2}{2} f + \frac{m^2 \psi^2}{4} + \lambda (3 \psi^2 + 3 H^2 + 2 \dot{H}) + (3\psi + 2 H ) \dot{\lambda} +\ddot{\lambda}.  {\label{13}}  
\end{align}

Using Eqs. (\ref{9}), (\ref{10}) and (\ref{11}), Eqs. (\ref{12}) and (\ref{13}) are simplified as,

\begin{align}
    \frac{1}{2} \rho =& \frac{k^2}{2} f + \frac{m^2 \psi^2}{4} + 3 \lambda (H^2 + \psi^2) - \frac{1}{2} m^2_{eff} H \psi, {\label{14}}\\
    \frac{1}{2} (p +\rho) =& - 2 \lambda \bigg( 1- \frac{m^2_{eff}}{12 \lambda} \bigg) \dot{H} + \frac{m^2_{eff}}{3}(H^2 + \psi^2 - 2 H \psi) + 2 k^2 f_Q \psi.{\label{15}}
\end{align}

We introduce a following set of dimensionless variables $ (\tau $, $ h $, $ \tilde{\rho} $, $ \tilde{\lambda} $, $ \Psi $, $ \tilde{Q} ) $ to simplified the field equations,

\begin{equation}{\label{16}}
    \tau = H_0 t,\hskip0.1in H= H_0 h, \hskip0.1in \rho = 6 k^2 H_0^2 \tilde{\rho}, \hskip0.1in \lambda = k^2 \tilde{\lambda}, \hskip 0.1in \Psi = H_0 \psi, \hskip 0.1in  Q = H_0^2 \tilde{Q}, \hskip 0.1in f = H_0^2 F.
\end{equation}

where $ H_0 $ represents the present value of the Hubble parameter.  The Eqs. (\ref{9}), (\ref{10}),  (\ref{14}) and (\ref{15}) are obtained as,

\begin{align}
     \frac{d \psi}{d \tau} =& \frac{d h}{d \tau} + 2 h^2 + \Psi^2 - 3 h \Psi, \label{17}\\
     \frac{d \tilde{\lambda}}{d \tau} =& - \bigg( \frac{M^2}{6} + 2 F_{\tilde{Q}} + 2 \tilde{\lambda} \bigg) \Psi = -\frac{1}{6} M^2_{eff} \Psi, \label{18}\\
     \frac{d h}{d \tau} =&  \frac{1}{1-M^2_{eff}/12 \tilde{\lambda}} \bigg( - \frac{3}{2} \gamma \frac{\tilde{\rho}}{\tilde{\lambda}} + \frac{\Psi}{\tilde{\lambda}} \frac{d F_{\tilde{Q}}}{d \tau} + \frac{M^2_{eff}}{6 \tilde{\lambda}} (h^2 + \Psi^2 - 2 h \Psi) \bigg),  \label{19} \\
     \tilde{\rho} =& \frac{1}{6} \bigg( F+ \frac{M^2 \Psi^2}{2} + 6 \tilde{\lambda} (h^2 + {\Psi}^2) - M^2_{eff} h \Psi \bigg). \label{20}
\end{align}
where
\begin{equation}{\label{21}}
  M^2_{eff} =  M^2 + 12 F_{\tilde{Q}} + 12 \tilde{\lambda} \hskip 0.2in with \hskip0.3in M^2 = \frac{m^2}{k^2}
\end{equation}
To solve the above field equations, we consider the following particular form of  $ f(Q) $ as $ f(Q) = ({H_0}^2) (\alpha_1 + \alpha_2 \hskip0.05in log ({H_0^{-2}} Q)) $  where $ \alpha_1 $ and $ \alpha_2 $ are arbitrary constants. So that, from Eq. \ref{16}, we get $ F(\tilde{Q}) = \alpha_1 + \alpha_2 log (\tilde{Q}) $ and $ F_{\tilde{Q}} = \frac{\alpha_2}{\tilde{Q}} = \frac{\alpha_2}{6 \Psi^2} $.
\\

By using the transformation $ \dot{z} = -(1+z)H $, the field Eqs. (\ref{17}), (\ref{18}), (\ref{19}), and (\ref{20}) are expressed in terms of red-shift $ z $ as follows:

\begin{align}
    -(1+z) h(z) \frac{d \Psi(z)}{d z} =& -(1+z) h(z) \frac{d h}{d z} + 2 h^2(z) + \Psi^2(z) - 3 h(z) \Psi(z), \label{22}\\
    (1+z) h(z) \frac{d \tilde{\lambda}}{d z} =& \frac{1}{6} M^2_{eff}(z) \Psi(z),  \label{23} \\
    -(1+z) h(z) \frac{d h(z)}{d z} =& \frac{1}{1- M^2_{eff}(z) /12 \tilde{\lambda}(z)} \bigg( - \frac{3}{2} \gamma \frac{\tilde{\rho}(z)}{\tilde{\lambda} (z)} + \frac{\Psi (z)}{\tilde{\lambda} (z)} (-(1+z) h(z)) \frac{d F_{\tilde{Q}}}{d z} + \frac{M^2_{eff}}{6 \tilde{\lambda}(z)} (h^2(z) + \Psi^2(z) - 2 h(z) \Psi(z)) \bigg), \label{24}\\
    \tilde{\rho}(z) =&  \frac{1}{6} \bigg( F+ \frac{M^2 \Psi^2(z)}{2} + 6 \tilde{\lambda} (z) (h^2(z) + \tilde{\lambda}^2(z)) - M^2_{eff}(z) h(z) \Psi(z) \bigg). \label{25}
\end{align}
where
\begin{equation}{\label{26}}
    M^2_{eff}(z) = M^2 +  2 \frac{\alpha_2}{ \Psi^2(z)} + 12 \tilde{\lambda}(z)
\end{equation}

\begin{figure}[h!]
\begin{center}
     \subfloat[]{\label{hz} \includegraphics[scale=0.40]{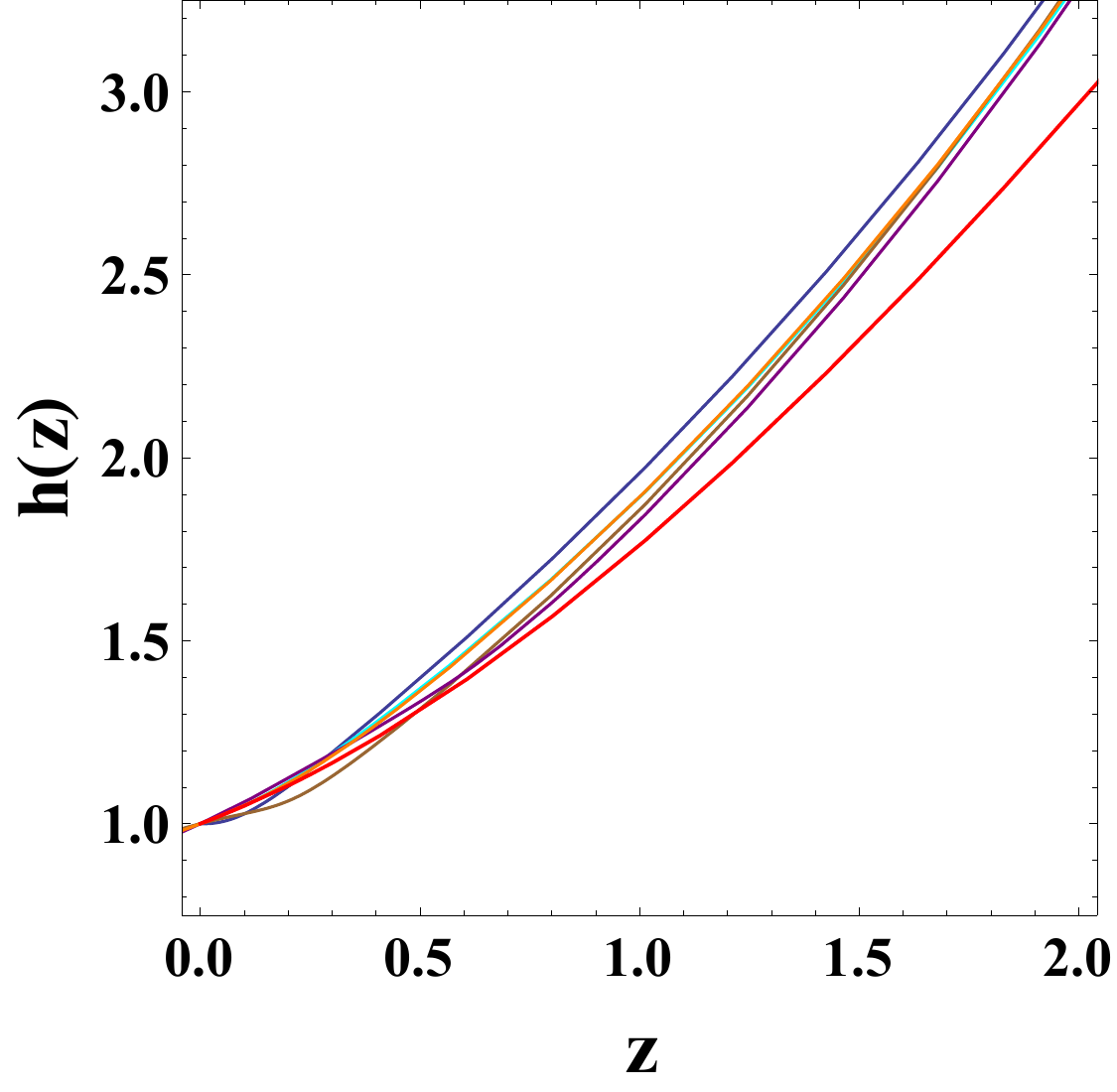}}\hfill
     \subfloat[]{\label{qz} \includegraphics[scale=0.40]{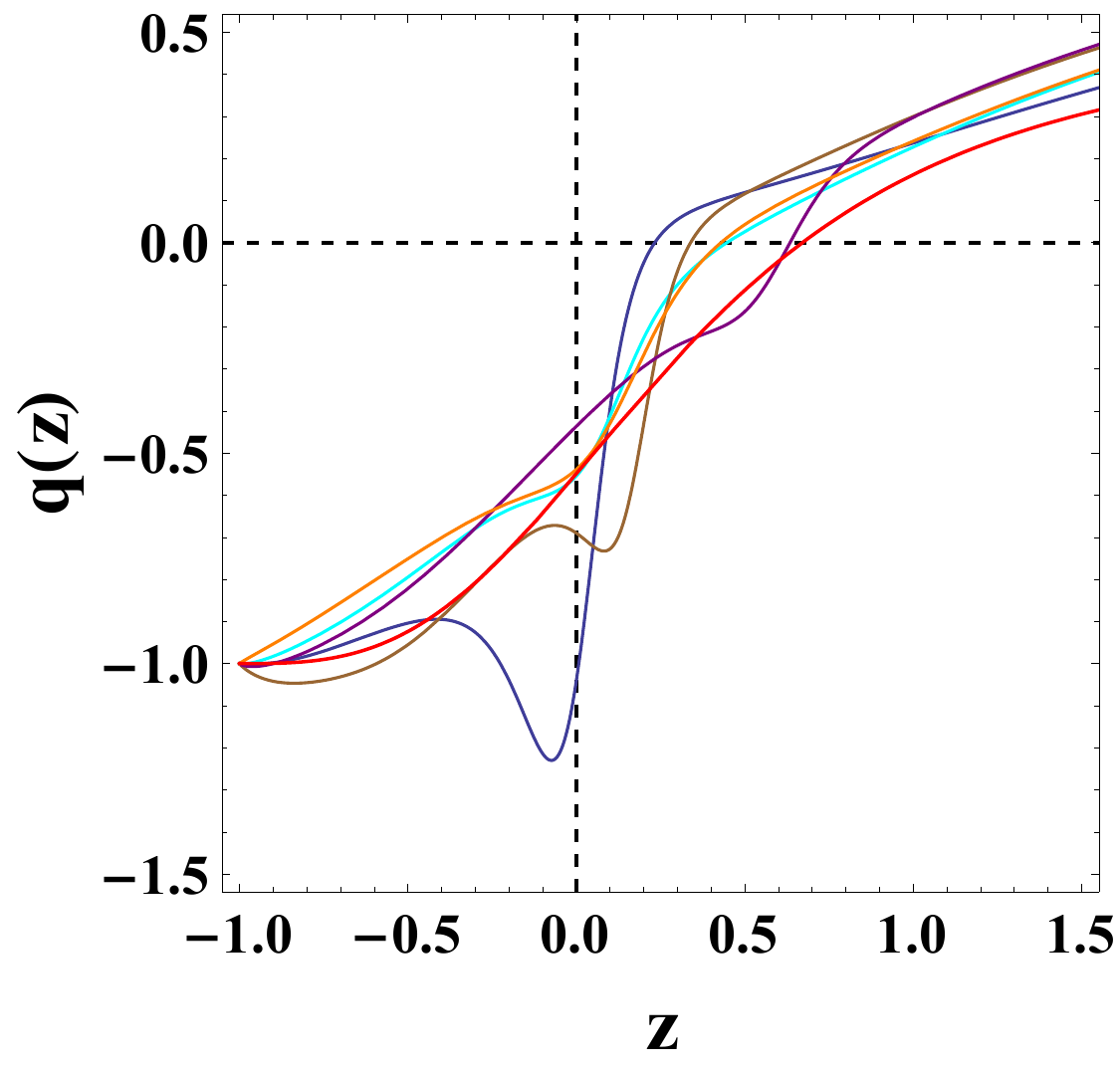}}
\end{center}
\caption{The evolution of the Hubble parameter and deceleration parameter over redshift $ z $ are described in the six plots in the Figs. (\ref{hz}) and (\ref{qz}) respectively. The five plots in each figure correspond to the five different set values of 3-tuple ( $ \alpha_1 $, $ \alpha_2 $, and the mass of the Weyl field $ M $) as $ (1,-1, 0.95) $ in Blue color, $ (-2.2,-5,5) $ in Cyan color, $ (2,-3,4) $ in Brown color, $ (-1,-3,4) $ in Purple color and $ (-1.5,-2.5,3) $ in Orange color. The sixth red-colored plot is that of the $ \Lambda $CDM model with the purpose of comparing our results with the standard model.}
\end{figure}

\begin{figure}[h!]
\begin{center}
     \subfloat[]{\label{lz} \includegraphics[scale=0.40]{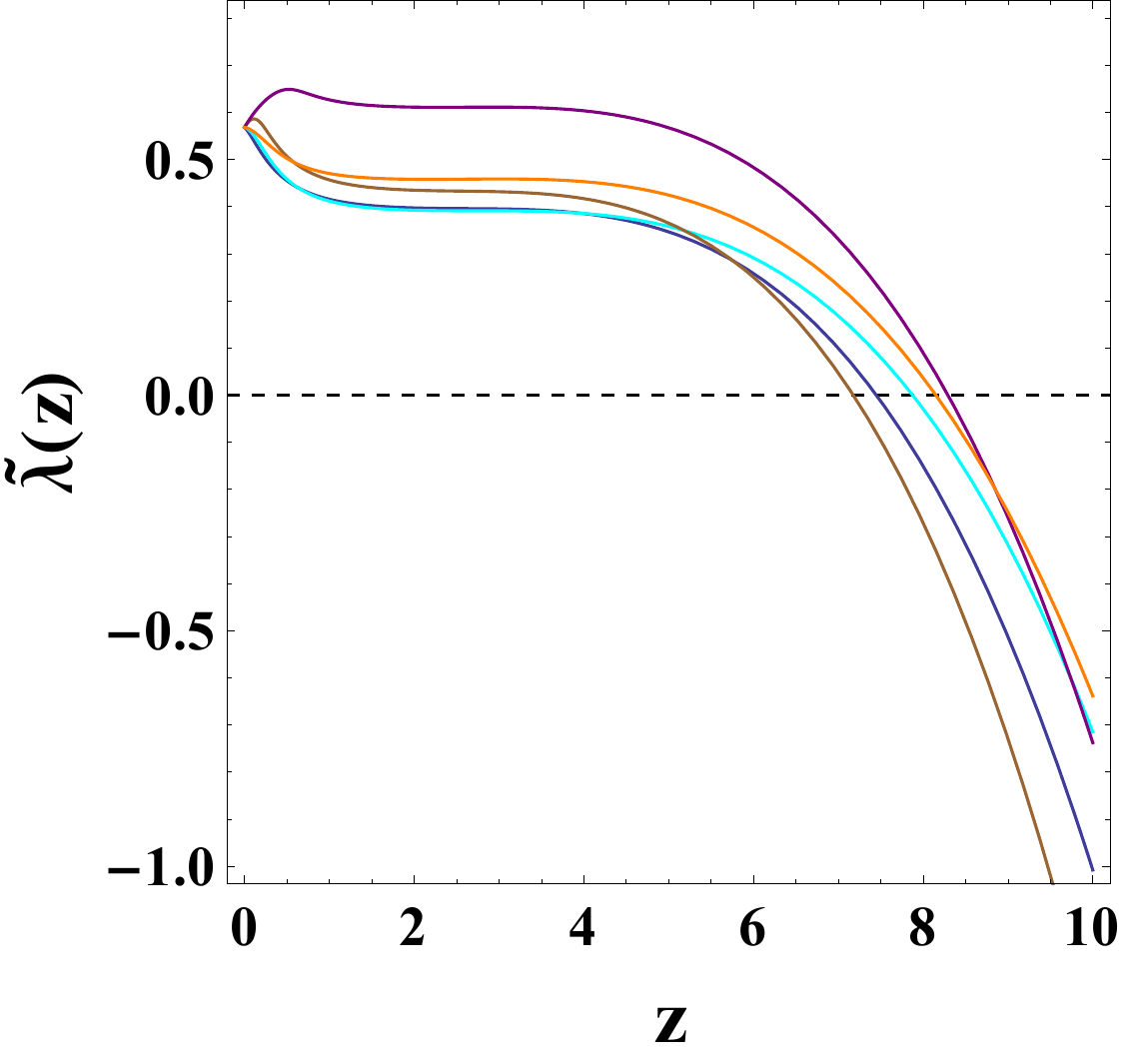}}\hfill
     \subfloat[]{\label{pz} \includegraphics[scale=0.40]{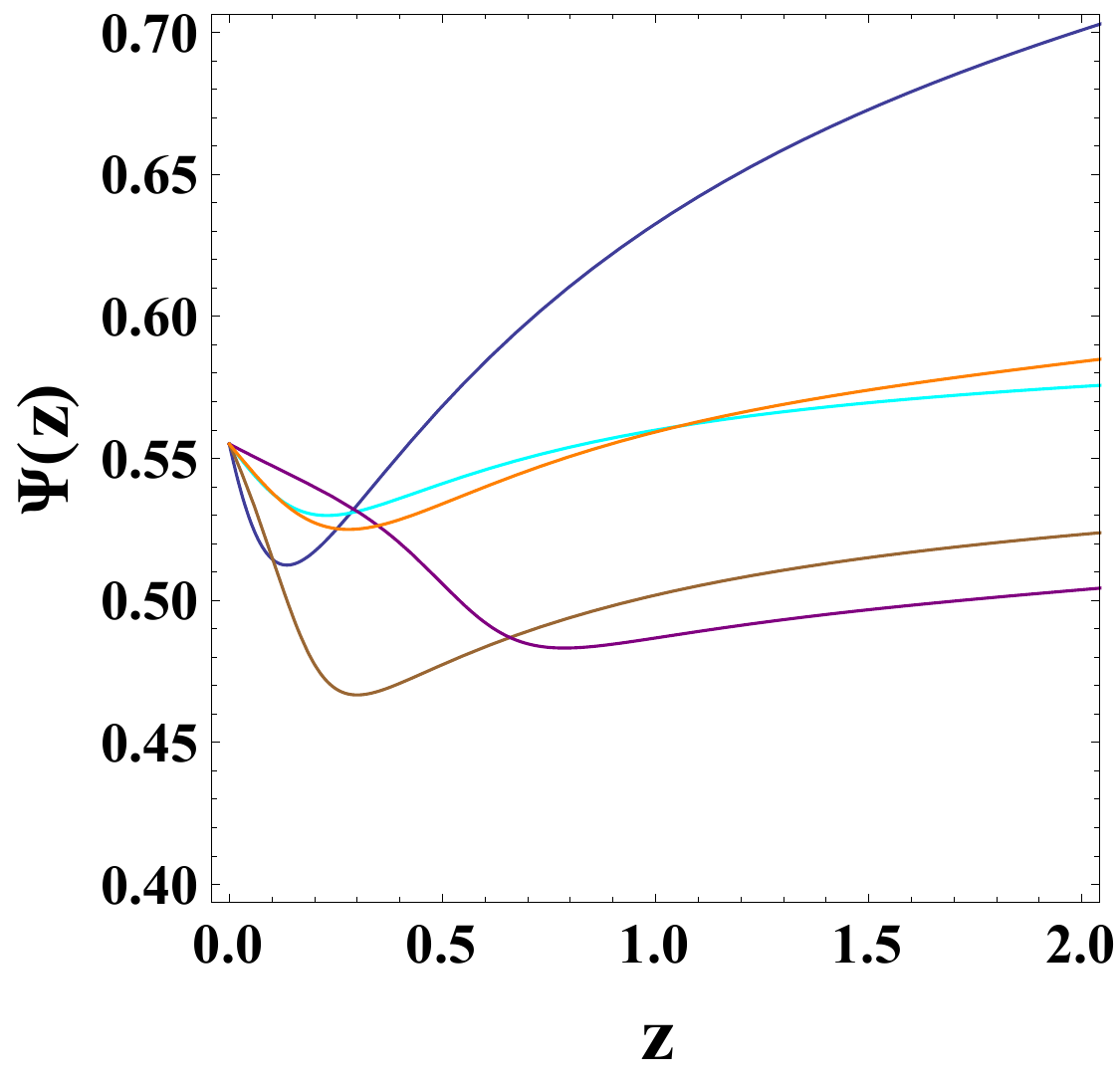}}
\end{center}
\caption{The plots of the  Lagrange multiplier ($ \tilde{\lambda} $) and Weyl vector ($ \Psi $) over red shift $ z $ . $ \Psi $ is associated with $ w $ as $ w^2 =  \frac{\Psi^2}{ H_0^2} $ and $\tilde{\lambda} = \frac{\lambda}{k^2} $.  The five plots in each figure correspond to the five different set values of 3-tuple ( $ \alpha_1 $, $ \alpha_2 $, and the mass of the Weyl field $ M $) as $ (1,-1, 0.95) $ in Blue color, $ (-2.2,-5,5) $ in Cyan color, $ (2,-3,4) $ in Brown color, $ (-1,-3,4) $ in Purple color and $ (-1.5,-2.5,3) $ in Orange color.}
\end{figure}

\begin{figure}[h!]
\begin{center}
    {\label{rho(z)}\includegraphics[scale=0.40]{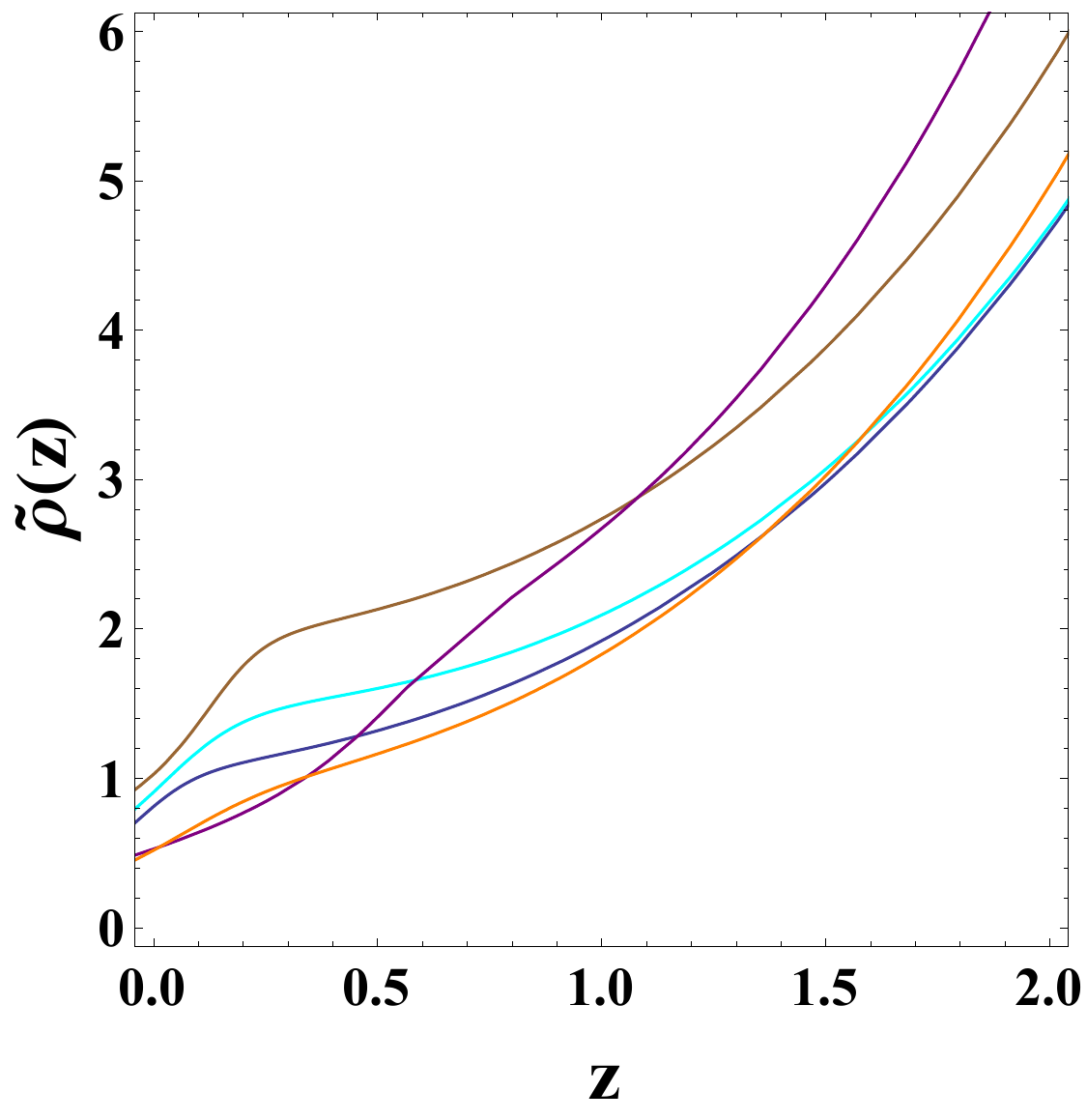}}
\end{center}
\caption{The plot of energy-matter density $ \tilde{\rho}(z) $ vs. redshift $ z $. The five plots in each figure correspond to the five different set values of 3-tuple ( $ \alpha_1 $, $ \alpha_2 $, and the mass of the Weyl field $ M $) as $ (1,-1, 0.95) $ in Blue color, $ (-2.2,-5,5) $ in Cyan color, $ (2,-3,4) $ in Brown color, $ (-1,-3,4) $ in Purple color and $ (-1.5,-2.5,3) $ in Orange color.}
\label{rhoz}
\end{figure}

We solve the above system of differential Eqs. (\ref{22})-(\ref{24}) numerically by taking the initial values $ h(0) = 1 $, $ \tilde{\lambda}(0) = 0.568 $ and $ \Psi(0) = 0.555 $. The numerical solutions of the Hubble parameter $ h(z) $, deceleration parameter $ q(z) $, Lagrange multiplier $ \tilde{\lambda}(z) $, Weyl vector $ \Psi(z) $ and the density parameter $ \rho $ are described and depicted in the form of plots in various Figs.  \ref{hz}, \ref{qz}, \ref{lz}, \ref{pz} and \ref{rhoz}. In each figure, we have presented five plots corresponding to the five different set values of 3-tuple ( $ \alpha_1 $, $ \alpha_2 $, and the mass of the Weyl field $ M $) as $ (1,-1, 0.95) $, $ (-2.2,-5,5) $, $ (2,-3,4) $, $ (-1,-3,4) $ and $ (-1.5,-2.5,3) $. 
\\

In Fig. \ref{hz}, it is observed that the Hubble parameter is monotonically increasing over redshift ($ z $) which means that it is decreasing over time ($ t $) in all the cases. It is also observed that our models are close to the standard $ \Lambda $CDM model initially for the redshift range between $ (0, 2) $. However, at higher redshift i.e. $ z > 2 $ there is a significant difference in the behavior of the growth of the Hubble parameter in our models and  $ \Lambda $CDM model.
We recall  the expression for the Hubble parameter $ H(z) $ and the deceleration parameter $ q(z) $ in the  $ \Lambda $CDM model as
\begin{equation}{\label{27}}
    H(z) = H_0 \sqrt{\Omega_{DM} (1+z)^3 + \Omega_\Lambda}
\end{equation}
and
\begin{equation}{\label{28}}
    q(z) = -1 + \frac{3(1+z)^3 (\Lambda_{DM})}{2(\Omega_\Lambda + \Omega_{DM} (1+z)^3))}
\end{equation}

where $ \Omega_{M} $ and $ \Omega_{\Lambda} $ are the density parameters of the cold dark matter (pressure less) and dark energy (also known as cosmological constant) respectively. The numerical values of density parameters are taken as $ \Omega_{DM} \equiv 0.3 $ and $ \Omega_{\Lambda} \equiv 0.7 $.
\\

The deceleration parameter ($ q $) in terms of Hubble parameter ($ H(z) $) and red-shift $ z $ is obtained as,
\begin{equation}{\label{29}}
     q(z) = (1+z) \frac{1}{H(z)} \frac{d H(z)}{dz} - 1
\end{equation}
Fig. \ref{qz} describes the evolution of the deceleration parameter $ q(z) $ for all the five values of model parameters ( $ \alpha_1 $, $ \alpha_2 $, and  $ M $). It is found that the deceleration parameter $ q(z) $  increases with a red shift ($ z $) and decreases with time ($ t $). We also observe that all the plots are found more or less nearer to the $ \Lambda $CDM model. There is a phase transition from deceleration in the past to acceleration at present. The value of the deceleration parameter at $ z=0 $ for different cases are $ -1.04 $, $ -0.55 $, $ -0.69 $, $ -0.44 $, and $ -0.54 $ approximately, and the corresponding transition redshifts are obtained as $ 0.2377 $, $ 0.4547 $, $ 0.3447 $, $ 0.635 $ and  $ 0.4333 $ (approximately).

Fig. \ref{lz} depicts the evolution of the Lagrangian multiplier $ \tilde{\lambda} $. It is observed that it decreases with redshift ($ z $) i.e. increases with time ($ t $). For the different values of model parameters ( $ \alpha_1 $, $ \alpha_2 $, and  $ M $), the graph behaves in a similar manner. However, $ \tilde{\lambda}(z) $ becomes negative approximately after $ z > 7 $.

Fig. \ref{pz} describes the evolution of the Weyl vector component $ \Psi(z) $ with respect to redshift ($ z $) for all the cases. It initially decreases then increases with higher values of redshift $ z $ in the range $ z \in (0,2) $. It becomes an increasing function after redshift $ z > 0.5 $.\\

Fig. \ref{rhoz} depicts the evolution of the matter density $ \tilde{\rho}(z) $. The matter density is monotonically increasing with the increasing values of redshift ($ z $) which means that it is decreasing with time ($ t $). However, the matter density entirely depends on the evolution of model parameters $ \alpha_1 $ and $ \alpha_2 $ and $ M $.

\newpage


\section{Observational Data Analysis}{\label{sec-4}}

\begin{figure}
\begin{center}
     \subfloat[]{\label{Herr} \includegraphics[scale=0.38]{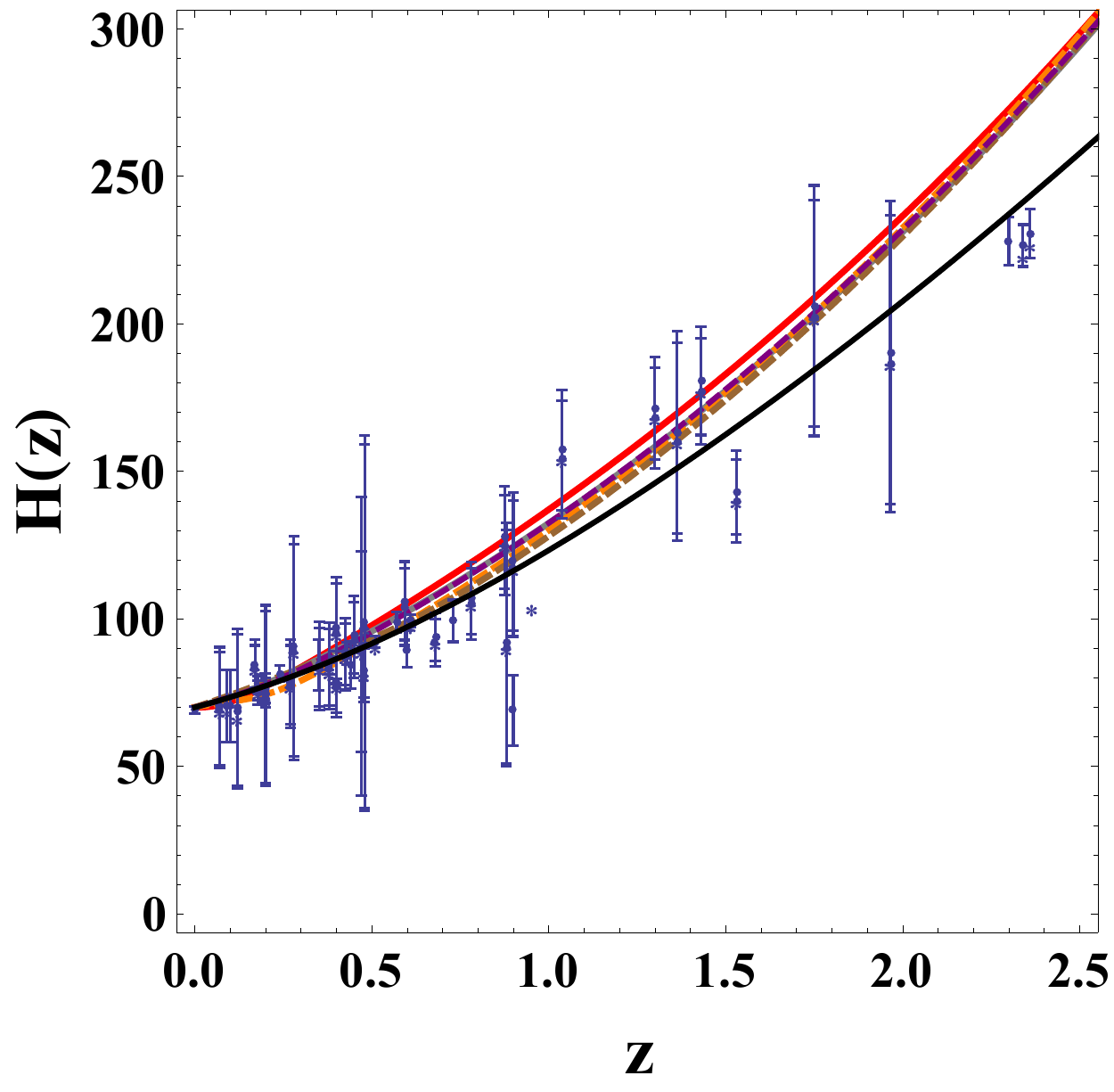}}\hfill
     \subfloat[]{\label{dl} \includegraphics[scale=0.40]{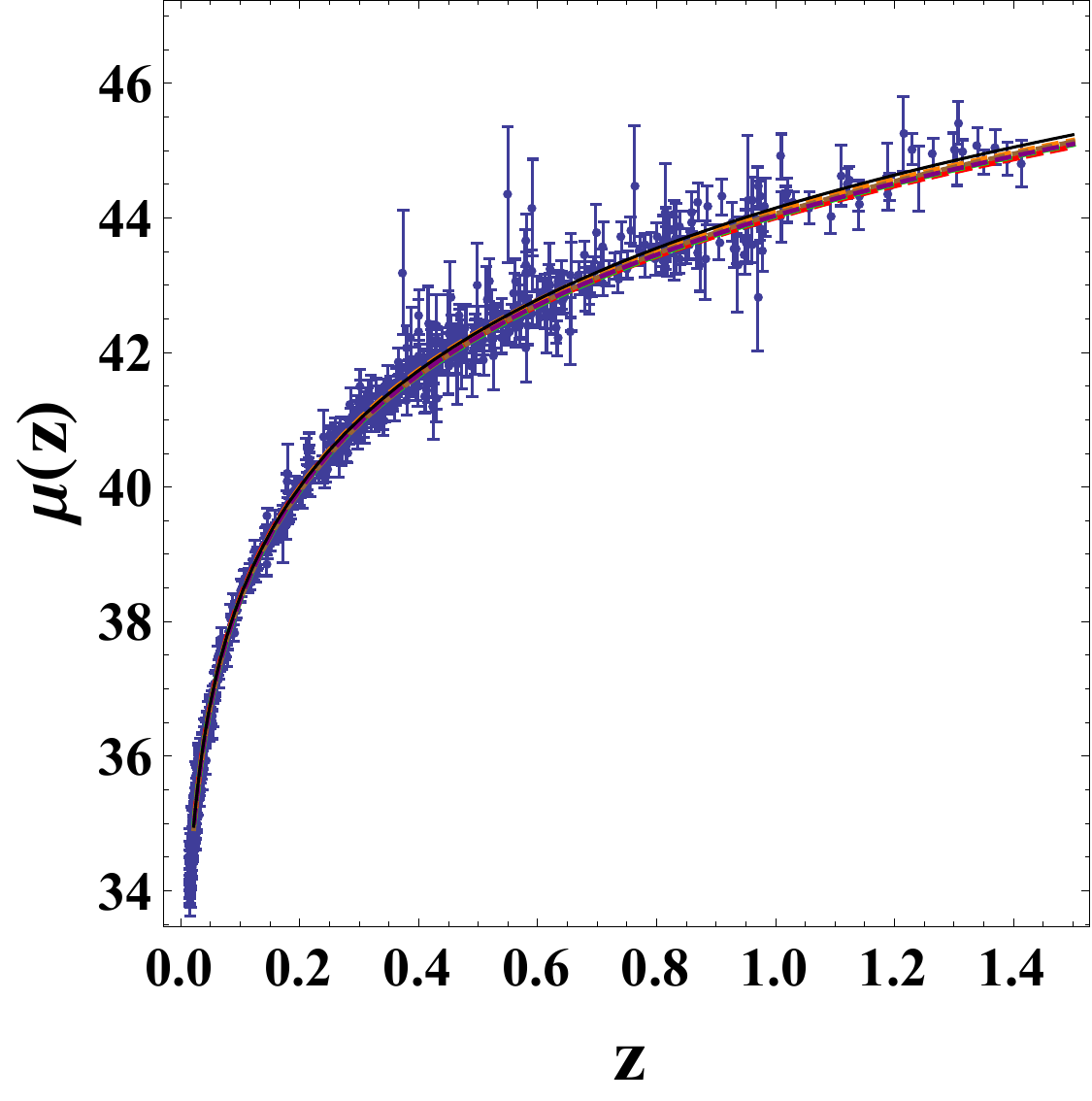}}
\end{center}
\caption{The two figures contain error bar plots of $ 77 $ Observational Hubble $ H(z) $ data set and $ 580 $ union 2.1 SN Ia Distance Modulus data set vs. redshift $ z $. The five regular plots in each figure are our theoretical plots of Hubble parameter and distance modulus corresponding to the five different set values of 3-tuple ( $ \alpha_1 $, $ \alpha_2 $, and  $ M $) as $ (1,-1, 0.95) $ in Blue color, $ (-2.2,-5,5) $ in Cyan color, $ (2,-3,4) $ in Brown color, $ (-1,-3,4) $ in Purple color and $ (-1.5,-2.5,3) $ in Orange color. The red colored plot represents the $ \Lambda $CDM model.}
\end{figure}

In this section, we use the three observed data sets namely the Observed Hubble data set of $ 77 $ points, the $ 580 $ distance modulus SNIa data set, and the $ 1048 $ supernova Pantheon data sets of apparent magnitudes to compare our theoretical results with those of observed data sets with the help of error bar plots. We have also computed the Chi-square to see the order of fit.\\
Fig. \ref{Herr} contains five theoretical plots of Hubble parameter $ H(z) $ corresponding to the five different set values of model parameters ( $ \alpha_1 $, $ \alpha_2 $, and  $ M $) and a red-colored plot corresponding to $ \Lambda $CDM model along with the observed  Hubble parameter data set points and corresponding error bars for different redshifts in the range ($ 0 \leq z \leq 2.5 $).  It is observed that our theoretical plots pass closely to the data set points as well as the  $ \Lambda $CDM plot. We also calculated the following Chi-square to see statistically the order of fit and we have found that $ \chi^2 =  77.908 $, $ 52.4227 $, $ 23.555 $, $ 31.6531 $ and $ 47.2343 $ respectively which is a good fit. 

\begin{equation}\label{30}
\chi^{2} = \sum\limits_{i=1}^{77}\frac{(H_{th}(z_{i}) - H_{ob}(z_{i}))^{2}}{\sigma {(z_{i})}^{2}},
\end{equation}
where $ H_{th}( H_0 * h_{th}) $ and $ H_{ob} $ are the theoretical and observational values of the Hubble parameter at redshift $ z $.  $ H_0 $ is the current value of the Hubble parameter and it is taken as $ 70  \hskip0.02in Mpc/sec/km $. 
\\   

The luminosity distance ($ d_L $) plays a very important role in astronomy as it determines the distance through the luminosity of a stellar object. The luminosity distance of any object is given by \cite{Copeland:2006wr}
\begin{equation}{\label{31}}
D_l(z) = (1+z)H_0 \int_0^z{\frac{1}{H(z*)} dz*},
\end{equation}
and the distance modulus of a luminous object is related to the luminosity distance through the following equation:
\begin{equation}\label{32}
\mu(z) = m_b - M = 5 Log D_l(z) + \mu_{0},
\end{equation}
where $ m_b $ and $ M $ are the apparent and absolute magnitude of the object and $ \mu_0 = 25 + 5 Log \big(\frac{c}{H_0} \big) $.\\
Fig. \ref{dl} contains five theoretical plots of Distance modulus $ \mu(z) $ corresponding to the five sets of values of model parameters ( $ \alpha_1 $, $ \alpha_2 $, and  $ M $) and a red colored plot corresponding to $\Lambda$ CDM model. It also carries $ 580 $ union 2.1 SNIa distance modulus data set points and error bars for different redshifts in the range ($ 0 \leq z \leq 1.5 $).  It is observed that our theoretical plots pass closely to the data set points as well as the  $\Lambda$ CDM plot.
We also calculate the following Chi-square to see statistically the order of fit and we have found that $ \chi^2 =  598.321 $, $ 589.545 $, $ 575.795 $, $ 585.698 $ and $ 585.227 $ respectively which is a good fit. 
\begin{equation}{\label{33}}
\chi_{\mu}^{2} = \sum\limits_{i=1}^{580}\frac{(\mu_{th}(z_{i}) - {\mu}_{ob}(z_{i}))^{2}}{\sigma {(z_{i})}^{2}},
\end{equation}
where $ \mu_{th} $  and $ \mu_{ob} $ are the theoretical and observational values of the distance modulus at redshift $ z $.

\begin{figure}
\begin{center}
    { \includegraphics[scale=0.38]{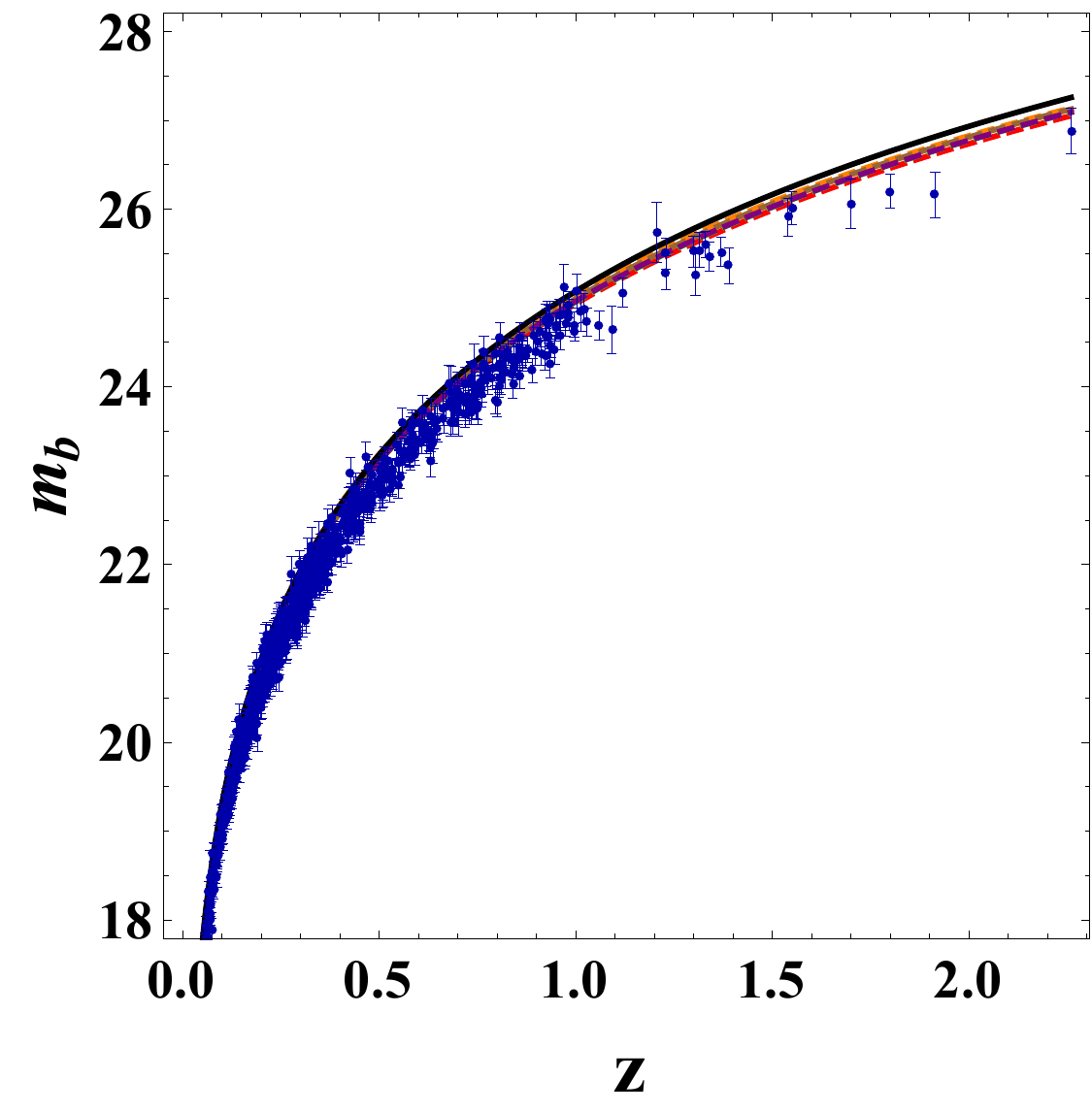}}
\end{center}
\caption{The figure contains error bar plots of 1048 pantheon data points of apparent magnitudes for redshifts in the range ($ 0 \leq z \leq 2.26 $). The five regular plots in the figure are our theoretical apparent magnitudes plots corresponding to the five different set values of 3-tuple ( $ \alpha_1 $, $ \alpha_2 $, and $ M $) as $ (1,-1, 0.95) $ in Blue color, $ (-2.2,-5,5) $ in Cyan color, $ (2,-3,4) $ in Brown color, $ (-1,-3,4) $ in Purple color and $ (-1.5,-2.5,3) $ in Orange color. The red color plot represents the $ \Lambda $CDM model}
\label{mb}
\end{figure}

The Apparent distance can be calculated from Eq. (\ref{32}),
\begin{equation}{\label{34}}
m_b = M + \mu(z) \hskip0.1in = -19.07 + \mu(z)
\end{equation}
where $ M $ is the absolute magnitude of the object and $ \mu $ is the distance modulus.
Fig. \ref{mb} contains five theoretical plots of apparent magnitude $ m_b(z) $ corresponding to the five sets of values of model parameters ( $ \alpha_1 $, $ \alpha_2 $, and  $ M $) and a red-colored plot corresponding to $\Lambda$ CDM model. Fig. also carries 1048 Pantheon data set points of apparent magnitudes and error bars for redshifts in the range ($ 0 \leq z \leq 2.26 $).  It is observed that our theoretical plots pass closely to the data set points as well as the  $\Lambda$ CDM plot.
We also calculate the following Chi-square to see statistically the order of fit and we have found that   $ \chi^2 =  5855.05 $, $ 5123.7 $, $ 6720.94 $, $ 4915.24 $ and $ 5181.11 $ respectively which is a good fit.

\begin{equation}{\label{35}}
\chi_{m_b}^{2} = \sum\limits_{i=1}^{1048}\frac{(m_{bth}(z_{i}) - m_{bob}(z_{i}))^{2}}{\sigma {(z_{i})}^{2}},
\end{equation}
where $ m_{bth} $  and $ m_{bob} $ are the theoretical and observational values of the distance modulus at redshift $ z $.

\section{Conclusion}{\label{sec-5}}
In this paper, we have explored an FLRW accelerating universe model in the Weyl type $ f(Q) $ gravity by taking the particular functional form of $ f(Q) $ as$ f(Q) = ({H_0}^2) (\alpha_1 + \alpha_2 \hskip0.05in log ({H_0^{-2}} Q)) $ . We solve the field equations numerically by taking the initial values of model parameters $ h(0) = 1 $, $ \tilde{\lambda}(0) = 0.568 $ and $ \Psi(0) = 0.555 $ and
 five different set values of 3-tuple parameters ( $ \alpha_1 $, $ \alpha_2 $, and the mass of the Weyl field $ M $) as $ (1,-1, 0.95) $, $ (-2.2,-5,5) $, $ (2,-3,4) $, $ (-1,-3,4) $ and $ (-1.5,-2.5,3) $. The numerical solutions of the Hubble parameter $ h(z) $, deceleration parameter $ q(z) $, Lagrange multiplier $ \tilde{\lambda}(z) $, Weyl vector $ \Psi(z) $ and the density parameter $ \rho $ are described and depicted in the form of plots in various figures  \ref{hz}, \ref{qz}, \ref{lz}, \ref{pz} and \ref{rhoz}. In each figure, we have presented five plots corresponding to the five different set values of  parameters ( $ \alpha_1 $, $ \alpha_2 $, and $ M $) . The salient features of the model are described in brief as follows:  
\begin{enumerate}
     \item The model shows a transition from decelerating in the past to acceleration at present which means that the deceleration parameter $ q $ was positive in the past and it is negative at present. The value of the deceleration parameter at $ z=0 $ for different cases are $ -1.04 $, $ -0.55 $, $ -0.69 $, $ -0.44 $, and $ -0.54 $ approximately, and the corresponding transition redshifts are obtained as $ 0.2377 $, $ 0.4547 $, $ 0.3447 $, $ 0.635 $ and  $ 0.4333 $ (approximately).  
     
     \item  We have solved Weyl type $ f(Q) $ gravity field equations numerically and have obtained numerical solutions to the Hubble and deceleration parameters, distance modulus, and apparent magnitudes of stellar objects like SNIa Supernovae. 
     \item We have also obtained numerical solutions for the Weyl vector ($ w $), non-metricity scalar ($ Q $), and the Lagrangian multiplier ($ \lambda $) appearing in the action of $ f(Q) $ gravity.
     \item We have compared the theoretical results of Hubble and deceleration parameters with those of the standard $ \Lambda $CDM model. From Fig. \ref{hz} and \ref{qz}, it is found that our models are coinciding with the standard $ \Lambda $CDM in the range of redshift $ z \in (0,2)$.
     \item In order to make our model compatible on observational grounds, we use three types of data sets: The Observed Hubble data set of $ 77 $ points, $ 580 $ distance modulus union 2.1 SNIa data set, and $ 1048 $ supernova Pantheon data sets of apparent magnitudes. We have compared our theoretical results with the error bar plots of the three data sets described earlier and it is found that our results fit well with the observed data set points. 
     \item The model envisages a unique feature that although the universe is filled with perfect fluid as dust whose pressure is zero, the weyl vector dominance $ f(Q) $ creates acceleration in it. 

\end{enumerate}

\section{Appendix: Weyl Geometry in Brief.}
The Riemann geometry permits parallel transportation of a vector along an infinitesimal loop in such a way that its magnitude remains constant whereas its direction may change as per the nature of the intrinsic property of curved space-time. We may see it as follows:
The variation of components of a vector $ v^i $ on parallel transportation is given as:
\begin{equation}\label{36}
	\delta v^i = v^{k} R_{k l j}^i  s^{l j}
\end{equation} 
where $ s^{ l j} $ is the area of the loop and $  R_{k l j}^i $ is the Riemannian curvature tensor. 
It can be verified that the infinitesimal change in the magnitude of the vector $ v^{k} $ on parallel displacement through the loop is nil. 
\begin{equation}\label{37}
	\delta ( g_{ij} v^i v^j)= 2 v^k v^j R_{jkl\eta}  s^{l \eta}=0
\end{equation}
Weyl introduced an intrinsic vector field $ w_i $ and a semi-metric connection $ {\tilde{\Gamma}^\alpha}_{i j} $ which is defined as
\begin{equation}{\label{38}}
    {\tilde{\Gamma}^\alpha}_{i j} \equiv \Gamma^\alpha_{i j} + g_{i j} w^\alpha - \delta^\alpha_i w_j - \delta^\alpha_j w_i  
\end{equation}
where $ \Gamma^\alpha_{i j} $ is the Christoffel symbol with respect to the metric $ g_{i j} $.
The semi-metric connection means that it has both metric and vector components.
The curvature tensor corresponding to the newly defined semi-metric tensor is denoted as $ \tilde{R}_{i j \alpha \beta}$. It has a both symmetric and an anti-symmetric part which is given by
\begin{equation}{\label{39}}
     \tilde{R}_{ij \alpha \beta} = \tilde{R}_{(ij)\alpha \beta} + \tilde{R}_{[i j] \alpha \beta},     
\end{equation}
where
\begin{equation}{\label{40}}
     \tilde{R}_{[i j] \alpha \beta} = R_{i j \alpha \beta} + 2 \nabla_{\alpha} w_{[i g_j] \beta} + 2 \nabla_{\beta} w_{[j g_i] \alpha} + 2 w_{\alpha} w_{[i g_j] \beta} + 2 w_{\beta} w_{[j g_i] \alpha}  - 2 w^2 g_{\alpha[i g_j] \beta},
\end{equation}
and
\begin{equation}{\label{41}}
     \tilde{R}_{(i j)\alpha \beta} = g_{i j} W_{\alpha \beta} 
\end{equation}
respectively, and
\begin{equation}{\label{42}}
    W_{i j} = \nabla_j w_i - \nabla_i w_j.
\end{equation}
In the Weyl geometry, the infinitesimal change in the magnitude of the vector $ v^i $ on parallel displacement through the loop is not zero.
\begin{equation}{\label{43}}
     \delta |v| =|v| W_{l \eta} s^{l \eta},
\end{equation}
where $ |v|^2 = v_i v^i $. In it, the divergence of the metric tensor is not zero under the semi-metric affine connection. We get the following expression for it
\begin{equation}{\label{44}}
     Q_{\alpha i j} \equiv  \tilde{\nabla}_\alpha g_{i j} = \partial_\alpha g_{i j} - {\tilde{\Gamma}^\eta}_{\alpha i} g_{\eta j} - {\tilde{\Gamma}^\eta}_{\alpha j} g _{\eta i} = 2 w_\alpha g _{i j}.
\end{equation}
We note that in the Riemannian geometry, the covariant derivative of the metric tensor is zero, i.e. $ \nabla_\alpha g_{ij} = 0 $. 

The tensor $ Q_{\alpha i j} $ is a three-indexed tensor. It can not be fully contracted with the help of metric tensor $ g_{ij} $ (only even order tensors can be contracted to scalar). So it is proposed an alternative non-metricity scalar $ Q $ is defined as follows. 

\begin{equation}{\label{45}}
    Q \equiv - g^{i j} \bigg( {L^\alpha}_{\beta i}  {L^\beta}_{j \alpha} - {L^\alpha}_{\beta \alpha} {L^\beta}_{i j} \bigg).
\end{equation}

where $  {L^\alpha}_{i j} $ is defined as 

\begin{equation}{\label{46}}
   {L^\alpha}_{i j} = - \frac{1}{2} g^{\alpha  \gamma} \bigg( Q_{i \gamma j} + Q_{j \gamma i} - Q_{\gamma i j} \bigg).
\end{equation}

From Eqs. (\ref{44}) -  (\ref{46}), we get the following important relation,

\begin{equation}{\label{47}}
      Q = - 6 w^2.
\end{equation}

\end{document}